\documentclass[12pt]{iopart}

\usepackage{graphicx}
\usepackage{bm}
\usepackage{units}

\newcommand\rs[1]{{\scriptscriptstyle\rm #1}}
\newcommand\pdag{{\vphantom\dagger}}
\renewcommand\Re{{\mathop{\rm Re}}}
\def\openone{\leavevmode\hbox{\small1\normalsize\kern-.33em1}}

\begin{document}

\title{Majorana fermions coupled to electromagnetic radiation}

\author{Christoph Ohm and Fabian Hassler}

\address{Institute for Quantum Information, RWTH
Aachen University, 52056 Aachen, Germany}
\ead{ohm@physik.rwth-aachen.de\\[2ex]
\normalfont  September 2013}

\begin{abstract}
We consider a voltage-biased Josephson junction between two nanowires hosting
Majorana zero modes which occur as topological protected zero-energy
excitations at the junction.  We show that two Majorana fermions localized at
the junction, even though being neutral particles, interact with the
electromagnetic field and generate coherent radiation similar to the
conventional Josephson radiation.  Within a semiclassical analysis of the
radiation field, we find that the optical phase gets locked to the
superconducting phase difference and that the radiation is emitted at half the
Josephson frequency.  In order to confirm the coherence of the radiation, we
study correlations of the radiation emitted by two spatially-separated
junctions in a d.c.-SQUID geometry taking into account decoherence due to
spontaneous state-switches as well as due to quasi-particle poisoning.
\end{abstract}


\section{Introduction} 
Ever since its discovery, superconductivity has been of great importance for
the understanding of quantum coherence.  A particular example is the Josephson
effect which in the most simple terms can be understood as a reactive current
which is driven by a gradient of the superconducting phase and thus
establishes the macroscopic coherence of the latter \cite{Josephson:1962uq}. In
fact, most of the physical phenomena related to superconducting tunnel
junctions are governed by the Josephson equations which have their microscopic
origin in a coherent transfer of Cooper pairs through the thin barrier.  These
equation predict that microwave radiation, also called \emph{Josephson
radiation}, is produced by a voltage biased tunnel
junction \cite{Josephson:1962uq, Lee:1971lr}. The microwave radiation is
emitted coherently at the Josephson frequency $\omega_{\rs J} = 2eV/\hbar$
which is given by twice the voltage bias $V$ that is applied across the
junction \cite{Yanson:1965fj,Langenberg:1965lr}; here and below, $e>0$ denotes
the elementary charge.  With superconducting-semiconducting hybrid devices, it
is possible to imprint superconducting correlations onto semiconducting
nano-devices like quantum dots or quantum wires \cite{De-Franceschi:2010qy}.
Recently, the potential for the emergence of Majorana fermions in such hybrid
systems has attracted a lot of interest in condensed matter
physics \cite{brouwer:12,franz:13}. With an appropriate tuning of the physical
parameters, Majorana zero modes are expected to appear as end states at the
chemical potential of the
superconductor \cite{Kitaev:2001kx,Alicea:2012fk,Beenakker:2013uq,
Leijnse:2012qy}. In particular, they are predicted to occur in one-dimensional
semiconducting quantum wires in proximity to a conventional \emph{s}-wave
superconductor subject to a moderate magnetic field \cite{Oreg:2010yq,
Lutchyn:2010fj}. In fact, recent experimental works seem to be in agreement
with these predictions \cite{Mourik:2012lr, Rokhinson:2012uq, Das:2012uq,
Deng:2012fk}. In contrast to conventional Josephson junctions, single electrons
can be transferred coherently in the presence of Majorana zero
modes \cite{fu:10} leading to a $4\pi$-periodic current-phase
relationship \cite{Kitaev:2001kx}.  Because of this, the Josephson effect is
dubbed \emph{fractional Josephson effect} \cite{Kitaev:2001kx, Fu:2009lr} and
its observation would provide a clear evidence of Majorana fermions.

The a.c.\ fractional Josephson effect was introduced in
\cite{Kwon:2004zr} and describes the case of voltage biased
Josephson junctions with fractional supercurrents, see also \cite{Pikulin:2012lr, San-Jose:2012fk, Houzet:2013lr} for detailed
discussion of the a.c.\ effect.  As a consequence, non-trivial Shapiro-steps
with doubled height in the current-voltage relation emerge providing a
signature of the Majorana fermions \cite{Dominguez:2012ys, Virtanen:rt,
Rokhinson:2012uq}. Regarding the interaction of Majorana zero modes with
electromagnetic fields, it has been recently shown that Majorana fermions can
be manipulated by means of microwave driving through weak coupling of
microwave radiation to the states of a two Majorana
wires \cite{Schmidt:2013fk}. Furthermore, in \cite{Ginossar:fk} the
influence of Majorana fermions on the photon coupling of a Majorana-transmon
qubit has been investigated.
%
\begin{figure}[t]
  \centering
  \includegraphics[width=0.6\textwidth]{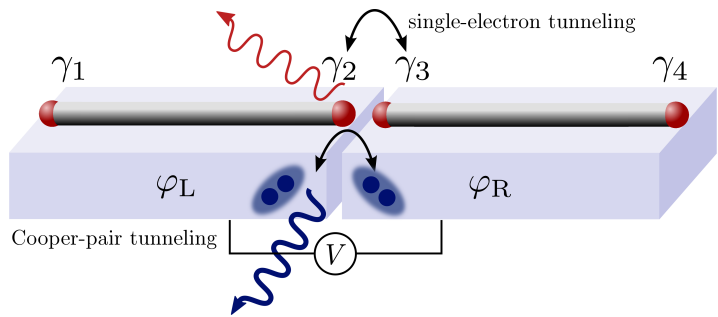} 
  \caption{Two bulk \emph{s}-wave superconductors (blue cuboids) forming a
  Josephson junction with superconducting phase difference $\varphi =
  \varphi_{\rs L} - \varphi_{\rs R}$.  On top of each of them, there is a
  semiconducting nanowire (gray cylinders) in the topological phase supporting
  two Majorana fermions (red spheres) at its ends.  We consider the case where
  the Josephson junction is voltage-biased such that Cooper pairs which move
  across the junction emit radiation at the Josephson frequency $2e V/\hbar$
  with $V$ the voltage applied.  This process is indicated by the bold wavy
  line.  Due to the presence of the Majorana fermions, there is the additional
  process allowed which proceeds via the tunneling of a single unpaired
  electrons together with the emission of radiation at at half the
  Josephson-frequency indicated by the thin wavy line.  In this paper, we will
  concentrate on the latter process.} \label{Fig:Maj-Jos-Junct}
\end{figure}

In this work, we want to focus on the radiation that arises from coupling
Majorana fermions to the electromagnetic field in a voltage-biased situation.
In analogy to the common a.c.\ Josephson effect we show that the
Majorana-induced Josephson radiation is coherent radiation with a frequency
that is half of the Josephson frequency.  Furthermore, there is a mutual
relationship between the emitted radiation and the state of the Majorana
fermions.  As a consequence, the system exhibits correlations between the
radiation fields emitted from different sources that are situated far away
from each other.  We will first introduce the coupling of the Majorana zero
modes with the electromagnetic field.  Then, we proceed outlining different
possible realizations of a voltage biased fractional Josephson junction.
Subsequently, we derive the semiclassical equation governing the radiation
field when the junction is placed in a cavity.  Moreover, we will discuss the
steady state solution and possible decoherence mechanisms.  We will finish
with the discussion of correlations of spatially separated radiation sources
and show that the superconducting coherence is partially imprinted onto the
radiation field.  In this respect, the system is closely related to prior work
\cite{Recher:2010qy,Godschalk:2011fk,Godschalk:2013lr,godschalk:13} which we
will discuss in details below.
%
\section{Josephson radiation from Majorana fermions}
%
\subsection{Dipole coupling of Majorana fermions}
In topological superconductors, Majorana fermions denoted by $\gamma_j$ appear
as quasi-particles in the middle of the gap.  Due to the build-in
particle-hole symmetry of superconductors in the mean field description these
solutions of the Bogoliubov-de Gennes equation are made-up from equal
superpositions of electrons and holes,
\begin{equation}
\gamma_j = \int \left[ w^*_j(\bm{r}) \psi(\bm{r}) + w_j(\bm{r})
\psi^{\dagger}(\bm{r}) \right] \, d^3 r,
\end{equation}
where $\psi(\bm{r})$ denotes the fermionic field operator of the electrons and
$w(\bm{r})$ is the wave function of a zero energy solutions of the
Bogoliubov-de Gennes equation which is localized at either end of the
nanowire \cite{Kitaev:2001kx,Oreg:2010yq,Lutchyn:2010fj}. The Majorana
operators obey the Clifford algebra $\{\gamma_j, \gamma_k\} =2 \delta_{jk}$.
In case the wire is manufactured on top of a Josephson junction additional
Majorana fermions have to be taken into account on either side of the
Josephson junction \cite{Fu:2009lr}. Accordingly, we consider four Majorana
bound states in the system denoted by $\gamma_1, \ldots, \gamma_4$, cf.\
figure~\ref{Fig:Maj-Jos-Junct}.  For the further analysis, it is convenient to
introduce two conventional fermionic operators
\begin{eqnarray}
f_{\rs L} &= \frac{1}{2} \left( \gamma_1 + i \gamma_2 \right), \qquad
f_{\rs R} = \frac{1}{2} \left( \gamma_3 + i \gamma_4 \right).
\end{eqnarray}
These fermionic operators account for the parity of the number of electrons on
either side of the Josephson junction with the parity given by $\mathcal{P}_x=
(-1)^{f^\dag_x f^\pdag_x} =\pm 1$, $x= \mathrm{L/R}$.  The fermionic
Hilbert-space is four dimensional and spanned by the vectors
$|\mathcal{P}_{\rs L}, \mathcal{P}_{\rs R}\rangle$.  It can be generated
from the ``vacuum'' state $|1,1\rangle$ via
\begin{eqnarray}
|1,1 \rangle, &|\bar1,\bar1\rangle = f^\dagger_{\rs L} f^\dagger_{\rs
R} 
|1,1\rangle, \qquad ({\rm even}), \nonumber\\
|\bar1,1\rangle = f^\dagger_{\rs L} \, |1,1\rangle,	 \qquad&|1,\bar1\rangle =
f^\dagger_{\rs R} |1,1\rangle, 	\qquad\;\;\;\,({\rm odd}),
\label{Eq:Majorana-states}
\end{eqnarray} 
where we have assembled the states according to the total parity
$\mathcal{P}=\mathcal{P}_{\rs L}\mathcal{P}_{\rs R}$ being even or odd and
introduced the shorthand notation $\bar 1= -1$.  It is important to notice
that the total parity is conserved such that we only have to consider two out
of the four states at one point.  Note that for an open system the parity
constraint may be violated, e.g., when taking quasi-particle poisoning into
account.

Due to the fact that the Majorana fermions are present, an exchange of single
electrons between the two sides of the Josephson junction becomes possible
which manifests itself in the current-phase relationship having a fundamental
period of $4\pi$.  Here, we are interested in the Josephson radiation emitted
by such a device.
For a conventional Josephson junction, the microscopic origin of the a.c.\ Josephson effect are
Cooper pairs tunneling across the junction thereby transferring a charge $-2e$
and emitting one photon at the Josephson frequency.  In the case of the
fractional Josephson effect, also single electron tunneling is allowed which
will lead to the emission of radiation at half the Josephson frequency.  
The coupling of the Majorana fermions to the electromagnetic field is provided
by a non-vanishing dipole matrix element entering the dipole Hamiltonian
\begin{equation}\label{Eq:Dipole-Operator}
	H_{\rm dip} = - \bm{d} \cdot \bm{E};
\end{equation}
here, $\bm{d}=-e \bm{r}$ is the dipole operator with $\bm{r}$ being the
position operator and $-e$ is the charge of the electron.  As explained in
details in \ref{Ap::BdG-Dipole}, the dipole operator of the Majorana zero
modes for the junction in figure~\ref{Fig:Maj-Jos-Junct} is given by 
\begin{equation}\label{eq:dipole}
\bm{d} =
-\frac{ie}2 \langle\bm{r} \rangle \cos(\varphi/2) \gamma_2 \gamma_3,
\end{equation}
with $\varphi= \varphi_{\rs L}- \varphi_{\rs R}$ the phase difference across
the junction and $ \langle\bm{r} \rangle= \langle w_2 | \bm{r } |w_3 \rangle$
the typical distance between the two Majorana fermions 2 and 3.  In deriving
(\ref{eq:dipole}), we have taken into account that only $\gamma_2$ and
$\gamma_3$ have a considerable overlap such that the dipole operator only
involves the Majorana fermions at the junction.  Note that even though the
Majorana fermions are charge neutral, the system exhibits a finite dipole
matrix element.  The reason is that even though the Majorana fermions carry
only information about the probability amplitude of chargeless quasiparticles,
the charge is provided by the superconducting condensate via the cosine term
involving the superconducting phase difference \cite{fu:10,heck:11}. If we
imagine that the junction is placed in a cavity supporting a single mode at
frequency $\omega$, the electrical field operator $\bm{E}$ can be written as
\begin{equation}
  \bm{E} \simeq 
  \sqrt\frac{\hbar\omega}{V} \bm{\epsilon} \left( a + a^\dagger \right)
\end{equation}
with $\bm{\epsilon}$ the polarization vector and $V$ the volume of the
cavity mode.  In conclusion, we have the Hamiltonian
\begin{equation}\label{Eq:Majorana-Dipole-Interaction}
H_{\rm dip} =	\frac{i g}{2}  \cos(\varphi/2) \gamma_{2} \gamma_{3} (a + a^\dagger)
\end{equation}
describing the interaction of the Majorana zero modes with the electromagnetic
radiation with $g \simeq e \sqrt{\hbar\omega/V} \, \bm{\epsilon} \cdot \langle
\bm{r} \rangle$ the light-matter interaction strength.\footnote{Recently,
light coupling to a transmon-like qubit systems involving Majorana fermions
has been discussed in \cite{Ginossar:fk}.  Different from us, the
dipole coupling discussed in that proposal has its origin in a capacitive
coupling between the superconducting islands.  } As the dipole operator is
oriented along the nanowire, we need to consider a mode with the electric
field having a component along the wire as otherwise the coupling $g$
vanishes.

Most importantly, the interaction Hamiltonian depends on half of the
superconducting phase difference $\varphi/2$ indicating that with each
tunneling event a single electrical charge is transferred across the junction.
Since the amount of a single electron charge $-e$ (half of a Cooper pair) is
transferred from one condensate to the other the dipole matrix element is
$4\pi$ periodic with respect to the superconducting phase difference.
Comparing (\ref{Eq:Majorana-Dipole-Interaction}) with the pure tunneling
contribution of the fractional Josephson effect \cite{Kitaev:2001kx,Fu:2009lr}
(with tunneling strength $w$)
\begin{equation}\label{Eq:Fractional-Josephson-effect}
 H_{\rm tun} = \frac{iw}{2}  \cos( \varphi/2 ) \gamma_2 \gamma_3
\end{equation}
shows that there is a close relationship between dipole and tunneling
interaction.  In both $H_{\rm dip}$ and $H_{\rm tun}$ the Majorana wave
functions need to have a considerable overlap such that single fermion
transfer is enabled.  Indeed both contributions
(\ref{Eq:Majorana-Dipole-Interaction}) and
(\ref{Eq:Fractional-Josephson-effect}) are complementary in the sense that
(\ref{Eq:Fractional-Josephson-effect}) is present at zero bias voltage and
(\ref{Eq:Majorana-Dipole-Interaction}) only becomes important for non-zero
bias as the dipole coupling allows for transitions at any energy difference
via emission/absorption of a photon carrying the energy surplus.  Accordingly,
the photon produced by a single electron transfer carries the energy $eV$
which corresponds to half of the Josephson frequency $\omega_{\rs J}/2$.
Therefore it makes sense to call the dipole coupling Hamiltonian
$H_{\rm dip}$ the a.c.\ analog of the d.c.\ Josephson effect.
%
\begin{figure}[t]
  \centering
  \includegraphics[width=0.6\textwidth]{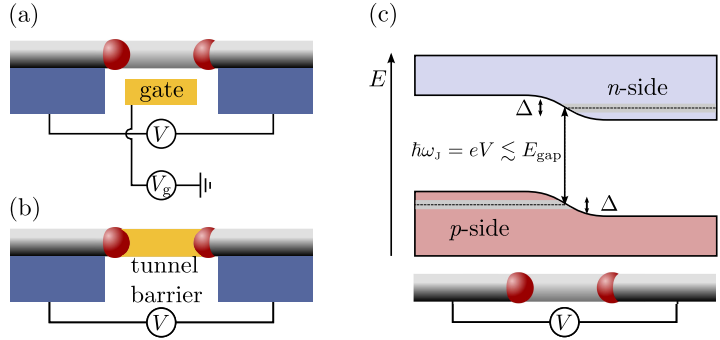}
  \caption{Possible ways to realize a fractional Josephson junction.  (a) A
  nanowire crossing the two superconductors which is capacitively coupled to a
  gate at voltage $V_{\rm g}$.  The gate controls the hybridization strength
  of the Majorana fermions at the Josephson junction.  The bias voltage $V$
  shifts the Fermi energies with respect to each other and leads to an a.c.
  Josephson effect.  The panel (b) shows a similar design with less tunability
  since the gated part of the nanowire is replaced with a narrow layer of
  insulator interrupting the semiconducting nanowire.  In both of these setups,
  the radiation frequency is bounded by the superconducting gap and thus
  typically is in the microwave regime.  This limitation can be overcome in
  the design (c) which shows the band diagram of a \emph{p-n} junction.  The
  different dopings lead to a pinning of the Fermi-energy in the valence band
  (left) and the conduction band (right).  Around the Fermi-energies a small
  superconducting gap is opened due to the proximity to the superconductors
  which leads additionally to two Majorana bound states (red spheres), on in
  the valence band and one in the conduction band.  The junction is operated
  in the forward bias regime with a potential difference comparable to the
  band-gap $E_{\rm gap}$.} \label{Fig:Junctions}
\end{figure}
%
\subsection{Possible realizations for fractional Josephson radiation}
As the main ingredient for the present proposal, a Josephson junction where
the wave function of the Majorana fermions on either side have sufficient
overlap is needed.  This junction then needs to be embedded in a cavity to
store and amplify the radiation field. Superconducting hybrid devices are very
flexible and offer many possibilities to obtain a desired physical effect.
Employing this freedom, we present three different designs for realizing
fractional Josephson radiation.

A possible setup involves a semiconducting nanowire covered by two
conventional \emph{s}-wave superconductors with a gateable junction in
between, see figure~\ref{Fig:Junctions}~(a).  Given the fact that the device
is in its topological phase, there are four Majorana fermions formed.  An
external bias voltage $V$ shifts the Fermi levels of wires with respect to
each other and the gate is used to control the Josephson coupling strength,
i.e., the size of the Majorana wave function overlap which enters the dipole
moment.  It is important to note that the bias voltage is bound by the
superconducting gap, $eV < |\Delta|$, as otherwise undesired quasiparticles
will be generated.  Given the typical size of a superconducting gap of the
order of a few Kelvin, the resulting Josephson radiation will be in the
microwave regime.  Obviously, the gated part of the nanowire can be replaced
by a narrow, insulating barrier included in the nanowire in the growth
process, cf.\ figure~\ref{Fig:Junctions}~(b).  Compared to the case (a)
discussed above, one would expect a larger overlap of the wave-function in
this case leading to a stronger light-matter interaction $g$ with the drawback
that the parameter is not tunable any more.

Inspired by \cite{Recher:2010qy}, there is another variant that makes use of a
superconducting \emph{p-n} diode, embedding a \emph{p-n} diode in a
semiconducting nanowire, see figure~\ref{Fig:Junctions}~(c).
 Doping \emph{p}- and \emph{n}-type carriers at the left and right side makes
a \emph{p-n} diode out of the wire and Majorana fermions are formed from the
valence and conduction band, respectively.  Recent investigations showed that
the topological phase of a semiconducting nanowire persists even in presence
of material imperfections (large dopant concentrations)
\cite{Adagideli:2013fr}.  The effect of dopants in the nanowire shifts the
chemical potential towards the conduction or valence bands depending on
carrier type and concentration.  Typically for semiconducting nanowires there
are two conduction bands and four hole bands, whereas the light-hole bands are
split-off because of the boundary conditions in quasi one-dimensional wires.
Light- and heavy-hole bands carry total angular momentum $J=\frac{3}{2}$ with
heavy holes being characterized by $m_J=\pm\frac{3}{2}$.  Out of the heavy
hole states, Majorana fermions are formed for the hole band exactly as for the
conventional electron band.  Across the \emph{p-n} junction the bands will
bend such that
 the electro-chemical potential remains constant.  By applying reversed bias
voltage, i.e., connect the \emph{p}-type side with the negative pole and the
\emph{n}-type side with the positive pole of a voltage source the junction
acts as an insulator up to voltages of the order of the band gap $E_{\rm
gap}$.  The corresponding energy diagram is shown in
figure~\ref{Fig:Junctions}~(c).  This setup allows to increase the Josephson
frequency from the superconducting gap to the band-gap $E_{\rm gap}$ of the
semiconductor.  For typical semiconductors with rather strong spin-orbit
interaction, as for instance InAs, the band gap is of the order of
$\unit[0.3]{eV}$ corresponding to wavelengths of around $\unit[4]{\mu m}$.
Hence, the radiation will be emitted in the mid-infrared regime.

Similar devices have been proposed in other system involving quantum dots and
a \emph{p-n} junction which allows to shift the Josephson frequency into the
optical frequency range \cite{Recher:2010qy, Godschalk:2011fk}.  The proposal
of \cite{Recher:2010qy} discusses incoherent radiation in the emission band
close to half the Josephson frequency and additionally coherent emission at
the Josephson frequency arising from the coherent transfer of Cooper pairs.
On the other hand, the proposal of \cite{Godschalk:2011fk} leads to coherent
radiation at half the Josephson frequency due to the fact that the device is
embedded in a microcavity; thus, the device has been named ``half-Josephson
laser''.  Although the optical phase of the laser is locked to the
superconducting phase difference, decoherence of a half-Josephson laser is
induced by spontaneous switches between different states of the quantum dot.
Further elaborations have demonstrated that coherence times of emitted light
from an array consisting of many emitters placed in a single cavity are
exponentially long \cite{Godschalk:2013lr,godschalk:13}.  The main difference
of these proposals to the present work is the fact that in our case there is
no need for quantum dots as the Majorana bound states are formed at the
interface without additional confinement.  Furthermore, the bound states are
automatically aligned with the chemical potential of the superconductor and
thus there is no need for (fine-)tuning of the dot parameters.
%
\section{Model for a Josephson junction and dynamics of the radiation}
%
\subsection{Model}\label{Subsec:Model}
Even though we have advertised three different physical implementation schemes
of Josephson radiation from Majorana fermions all of them can be described by
a single effective model.  This is because in all three setups the physical
process leading to radiation is tunneling of single electrons accompanied by
the emission of a photon as described by
(\ref{Eq:Majorana-Dipole-Interaction}).  Taking additionally the energy of
the photons in the cavity as well as possible overlap terms between Majorana
fermions on each wire section into account, we arrive at the model Hamiltonian
	\begin{equation}
    \fl
		H	= 	\frac{\varepsilon_{\rs L}}{2} \left(1 - i \gamma_{1} \gamma_{2} \right) 
			+ 	\frac{\varepsilon_{\rs R}}{2} \left(1 - i \gamma_{3} \gamma_{4}
      \right) + \hbar \omega a^\dagger a 	
      +	\frac{ig}{2} \cos[ \varphi(t)/2 ] \gamma_2 \gamma_3 (a +
      a^\dagger )		\label{Eq:Model-Hamiltonian}
	\end{equation}
which is the basis for the subsequent analysis; here, the overlap amplitudes
are denoted by $\varepsilon_{\rs L}$ and $\varepsilon_{\rs R}$ for Majorana
modes $\gamma_1, \gamma_2$ in the left and $\gamma_3, \gamma_4$ in the right
section of the nanowire, respectively.  These coefficients are decaying
exponentially with the separation $\varepsilon_{\rs{L/R}} \propto
e^{-L_{\rs{L/R}}/\xi}$, where $L_{\rs{L/R}}$ is the length of a particular
wire section and $\xi$ is the superconducting coherence length
\cite{Kitaev:2001kx}.  Furthermore, it has been assumed that Majorana modes
$\gamma_1$ and $\gamma_4$ are located too far away from each other such that
their overlap can be neglected compared to $\varepsilon_{\rs L},
\varepsilon_{\rs R}$.  The applied bias voltage is taken into account by a
time-dependent superconducting phase $\varphi(t) = \varphi(0) + 2eVt/\hbar$.
In deriving (\ref{Eq:Model-Hamiltonian}), we have neglected all quasiparticle
excitation above the superconducting gap which leads to the requirement that
$g |\langle a^\dagger + a \rangle |\ll |\Delta|$.  Moreover, we have used the
fact that we are interested in a situation where the photon frequency is
$\omega \approx \omega_J/2$.  Because of this, we neglected both the
contribution from the d.c.\ fractional Josephson effect
(\ref{Eq:Fractional-Josephson-effect}) as well as from usual Cooper pair
tunneling as those contributions are off resonance.  The model Hamiltonian
(\ref{Eq:Model-Hamiltonian}) describes the dynamics of the Majorana bound
states in a Josephson junction as well as dynamics of the radiation field.
Note that the superconducting condensate acts as a driving force in this
model.  It is convenient to represent (\ref{Eq:Model-Hamiltonian}) in terms of
a Fockspace basis consisting of states $|n\rangle \otimes |N\rangle$ where $|
N \rangle$ is the state of the photon mode occupied with $N$ photons and
$|n\rangle$ describing the fermionic states as introduced in
(\ref{Eq:Majorana-states}).

Next, we assume that the voltage $ eV =\hbar\omega_{\rs J}/2$ is large
compared to the other characteristic energies of the system $
\varepsilon_{\rs{L/R}}, g, \Omega = \omega - \omega_{\rs J}/2$.  We change via
the unitary transformation $U = \exp\left( - \frac{i}{2} \omega_{\rs J} t \,
a^\dagger a \right)$ from (\ref{Eq:Model-Hamiltonian}) into a rotating
frame where we neglect the rapidly oscillating terms $\propto \exp(-i
\omega_J/2)$. We then end up with the Hamiltonian
\begin{equation}
	H_{\rs{RWA}} = 		\frac{\varepsilon_{\rs L}}{2} \left(1 - i \gamma_{1} \gamma_{2} \right) 
			+ 	\frac{\varepsilon_{\rs R}}{2} \left(1 - i \gamma_{3} \gamma_{4} \right) 
			+ \hbar \Omega b^\dagger b  
			+	\frac{ig}{4} \gamma_2 \gamma_3 (b + b^\dagger ). \label{Eq:RWA-Hamiltonian}
\end{equation}
in the rotating wave approximation (RWA); here, we have introduced the new
operators $b = e^{ i \varphi(0)/2} \, a$ absorbing the
superconducting phase at the initial time $t=0$.
%
%
\subsection{Semiclassical approximation}\label{Subsec:Semiclassics}
As we are interested in the resonant regime of small $\Omega$, many photons
will accumulate in the cavity due to the driving of the superconducting phase.
If the cavity losses are small compared to the pumping rate, many photons will
accumulate in the cavity and the radiation field will approach a classical state
with a photon number $N=\langle b^\dag b \rangle \gg 1$.  We want to refer to
this limit as the semiclassical limit, because the degrees of freedom of the
nanowire are still treated quantum mechanically.  The semiclassical
approximation in (\ref{Eq:RWA-Hamiltonian}) amounts to replacing $b$ with the
complex number $\lambda = \langle b \rangle$.  The Hamiltonian $H_{\rs {RWA}}$
becomes a $4\times4$ matrix representing the Majorana bound states coupled to
a classical field $\lambda(t)$ which is still a dynamical variable of the
problem.  Even though $N$ is expected to becomes large, we still want to
demand $g\,|\Re \lambda| \ll |\Delta|$ to avoid interaction with the states
above the superconducting gap $\Delta$.  The equations of motion for
$\lambda(t)$ can be derived from the Heisenberg equation of motion for $b$.
This yields the differential equation for the classical radiation field
\begin{equation}\label{Eq:Adiabatic-equations-of-Motion}
\dot\lambda = -	\left( i \Omega + \Gamma \right) \lambda +
\frac{g}{2\hbar}  \langle \psi(t) | \gamma_2 \gamma_3 |\psi(t) \rangle,
\end{equation}
where we have introduced the cavity loss rate $\Gamma$ and the
(time-dependent) electronic state $|\psi(t)\rangle$. The state of the nanowire
evolves according to
\begin{equation}\label{eq:nw}
|\psi(t) \rangle = \mathcal{T} \exp \left( -\frac{i}{\hbar} \int\limits_0^t
H_{\rs{W}}( \lambda) dt' \right) |\psi(0)\rangle
\end{equation}
with $\mathcal{T}$ the time-ordering operator and 
\begin{equation}
H_\rs{W}(\lambda) =\frac{
\varepsilon_{\rs L}}2 (1- i \gamma_1 \gamma_2) +\frac{\varepsilon_{\rs R}}2
(1- i
\gamma_3 \gamma_4) + \frac{i g}2 \gamma_2 \gamma_3 \,\Re \lambda
\end{equation}
the Hamiltonian of
the nanowire driven by the time-dependent field $\lambda(t)$.  The two
equations (\ref{Eq:Adiabatic-equations-of-Motion}) and (\ref{eq:nw}) need to
be solved self-consistently.  As we are interested in the long-time dynamics
where the system approaches a stationary state with $\dot \lambda =0$,
we may assume that the radiation field $\lambda$ changes slowly in time and
the system adjusts adiabatically.  In this regime, the system remains in its
instantaneous eigenstates and $|\psi;\lambda \rangle$ which we will  assume in
the following.

\begin{figure}[t]
  \centering
  \includegraphics[width=0.45\textwidth]{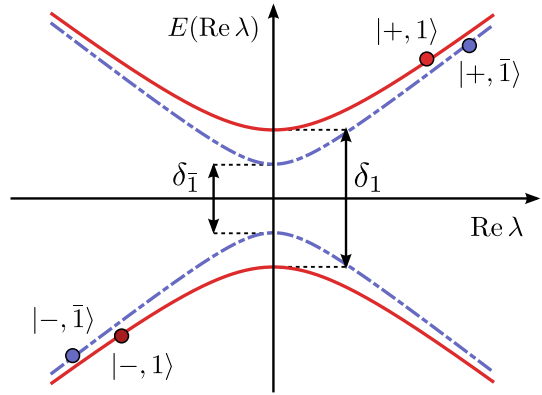}
  \caption{Spectrum of the electronic system as function of $\Re\lambda$.
  The spectrum has two distinct parity sectors indicated by solid (even) and
  dashed (odd) spectral-lines differing by their level splitting at zero
  field, $\delta_{\mathcal{P}}=\varepsilon_{\rs L} + \mathcal{P}
  \varepsilon_{\rs R}$.  During the initial time evolution, the radiation
  field drives the system several times through the avoided crossing until it
  approaches a stationary state.  There is exactly one stationary state of the
  radiation field, indicated by the dots, associated with each eigenstate of
  the electronic system.  }
  \label{Fig:SemiclassicalSpectrum}
\end{figure}

There are four instantaneous eigenstates of $H_\rs{W}$ which we will denote by
$|\pm , \mathcal{P}; \lambda \rangle$ with $\mathcal{P}=\pm 1$ labeling the
even and odd parity sectors and $\pm$ indicating whether the system is in the
upper or lower state respectively.  The corresponding eigenenergies are given
by
\begin{equation}
E_{\pm, \mathcal{P}} = \frac{1}{2} \Bigl( \varepsilon_{\rs L} +
\varepsilon_{\rs R} \pm \sqrt{g^2 \Re^2 \lambda + \delta_{\mathcal{P}}^2 }
\Bigr);
\end{equation}
here, $\delta_{\mathcal{P}} = \varepsilon_{\rs L} + \mathcal{P}\,
\varepsilon_{\rs R}$ is the size of the avoided crossing, see
figure~\ref{Fig:SemiclassicalSpectrum}.  The validity of the adiabatic
approximation is given by $|\langle +, \mathcal{P} |\gamma_2 \gamma_3 |
-,\mathcal{P} \rangle | \ll \frac{2}{\hbar} g\,|\Re \lambda| / |\frac{d}{dt}
\ln(\Re \lambda)|$ which using (\ref{Eq:Adiabatic-equations-of-Motion})
translates to the requirement $\hbar \sqrt{ \Omega^2 + \Gamma^2} \ll g \,|\Re
\lambda|$.

The initial dynamics of the radiation field before becomes close to the
stationary state is highly non-adiabatic and the electronic system will switch
many times between different eigenstates due to the driving of cavity field.
However, we are not interested in the transient dynamics and concentrate on
the stationary state of the radiation field which will turn out to be phase
locked to the superconducting phase difference, see below.  In the adiabatic,
regime the matrix elements needed in (\ref{Eq:Adiabatic-equations-of-Motion})
can be evaluated explicitly,
\begin{equation}
	\langle \pm, \mathcal{P} |\gamma_2 \gamma_3 |\pm, \mathcal{P} \rangle  
	= \mp i \frac{ g\,\Re\lambda}{\sqrt{ g^2\,\Re^2\lambda +
  \delta_\mathcal{P}^2}} ,
\end{equation}
assuming that the electronic system is in the state $|\psi\rangle =
|\pm,\mathcal{P}\rangle$.  Plugging the matrix element into the equation of
motion (\ref{Eq:Adiabatic-equations-of-Motion}), it becomes a non-linear
differential equations for the field amplitude $\lambda$. We can see that
apart from the trivial stationary solution $\lambda =0$, there is for each
eigenstate the second stationary solution
\begin{eqnarray}
\lambda_{\pm, \mathcal{P}} &= \pm \frac{1}{2} \, \sqrt{\frac{g^2}{4 \hbar^2 \, (\Omega^2 + \Gamma^2)} 
								- 4 \frac{\Omega^2 + \Gamma^2}{g^2} \, \left( \frac{\delta_{\mathcal{P}}}{\Omega}\right)^2 } 
                e^{i \arctan(\Gamma/\Omega)} \nonumber\\
							&\approx	\pm \frac{g}{4 \hbar \, \sqrt{\Omega^2 + \Gamma^2}} 
              e^{i \arctan(\Gamma/\Omega)}. \label{Eq:Stationary-Points}
\end{eqnarray} 	
fulfilling $\dot\lambda_{\pm, \mathcal{P}}=0$.  The nontrivial states are
stable (and correspondingly the trivial states unstable) if
$4|\delta_{\mathcal{P}}|/\Omega \leq g^2/ \hbar(\Omega^2 + \Gamma^2)$ which
is why we have neglected $\delta_\mathcal{P}$ by passing from the first to the
second line in (\ref{Eq:Stationary-Points}).  According to the balance
between driven pumping and cavity losses, the stationary field amplitude
depends on the cavity quantities $\Omega,\Gamma$ and the coupling $g$.  Going
back to the original non-rotating frame the stationary field configuration
describes an oscillating field $\langle a \rangle = \exp[-i
\omega_{\rs J}t/2 - i\varphi(0)/2 ] \lambda_{\pm, \mathcal{P}}$ at
frequency $\omega_{\rs J}/2$, i.e., half of the Josephson frequency.
Regarding the phase of the radiation field $\langle a \rangle$, we see that it
is \emph{locked} to half of the superconducting phase difference
$-\varphi(0)/2$ with an additional phase shift $\arctan(\Gamma / \Omega) $ due
to the cavity.  Moreover, the sign of the radiation field depends on the fact
whether the system is in the upper or lower state but (almost) not on
$\mathcal{P}$.  This locking mechanisms protects the emitted radiation from
diffusion of the phase as it is the case for conventional
lasers \cite{Scully:1967fk}. This is a consequence of the broken $U(1)$ phase
symmetry of the superconductors that is imprinted in the phase of the
radiation field.  we expect the coherence time of the emitted radiation to be
rather long as experiments measuring the relaxation of the persistent current
in superconductors have shown that the superconducting coherence lasts for
years \cite{Tinkham:1996fk}. As we show in the next section, these extremely
long coherence times can not be reached in realistic systems due to
spontaneous switches of the electronic system as well as quasiparticle
poisoning.
%
%
\section{Coherence of the radiation}\label{Sec:Coherence}

Switches of the electronic state of the wire are caused by two mechanisms,
emission of off-resonant photons and quasiparticle poisoning, that are present
even if the stationary state has been reached.  Whenever a spontaneous
switching event happens, the cavity field will be driven out of its stationary
state and eventually will approach another one.  Here, we will assume that
switches happen instantaneously on the time-scale $\Gamma^{-1}$ of the cavity.
While the system approaches another stable amplitude the time evolution may
get non-adiabatic again including many other switching processes as described
in Sec.~\ref{Subsec:Model}.  As noted above, in case the switching process
changes the electronic system from the upper to the lower branch (or the other
way round), it is accompanied with a change of $\pi$ of the phase of the
radiation field.  We will discuss two mechanism which lead to decoherence: The
first process we will discuss conserves the total parity $\mathcal{P}$.  By
emitting a non-resonant photon that carries the energy of the level splitting
$\hbar \tilde{\omega} \simeq g\,|\Re\lambda|$, a transition between those upper
and lower states at fixed parity is possible.  The transition rate for such a
process can be evaluated by Fermi's golden rule.  Using the fact that the
density of state of the photons in the cavity is Lorentzian, we obtain the
approximate transition rate \cite{Godschalk:2011fk}
\begin{equation}
	\Gamma_{\rs{F}} = \frac{\Gamma}{8} \left( \frac{\delta_{\mathcal{P}}}{g\,\Re^2\lambda} \right)^2,
\end{equation}
where we have assumed that the photon frequency is far detuned,
$\tilde\omega\gg \Gamma, \Omega$.  Given the fact that $\delta_\mathcal{P}$
depends exponentially on the separation of the Majorana fermions, we expect
that $\Gamma_\rs{F}$ will most likely be not the dominating process generating
decoherence but rather the fact that superconducting devices suffer from
\emph{quasi-particle poisoning}.  As soon as an extra quasi-particles tunnels
on one of the bulk superconductors the parity of the device is changed.
Additionally, also the upper state might be transformed into the lower state
as the quasiparticle changes $\mathcal{P}_{\rs L}$ or $\mathcal{P}_{\rs R}$
which do not commute with the Hamiltonian $H_\rs{W}$.  For concreteness, we assume
that a quasiparticle tunneling switches the state $|+,\mathcal{P}\rangle$ to
the states $|\pm,\bar{\mathcal{P}}\rangle$ with equal probability.
Measurements of transmon-type charge qubits have shown that quasi-particle
tunneling appears to happen on rather long time scales in the range from
microseconds up to milliseconds \cite{Riste:2013fk}.  In the following
quasi-particle tunneling will be modeled simply by a $\Gamma_{\rs{QP}}$.  On
top of these spontaneous switches, there are coherent Rabi oscillations with
frequency $g\,|\Re\lambda|$ taking place because of the influence of
counter-rotating terms that have been neglected in the RWA.  See
\ref{Ap:Counter-Rotating-Terms} for details.
%
\begin{figure}[t]
  \centering
  \includegraphics[width=\textwidth]{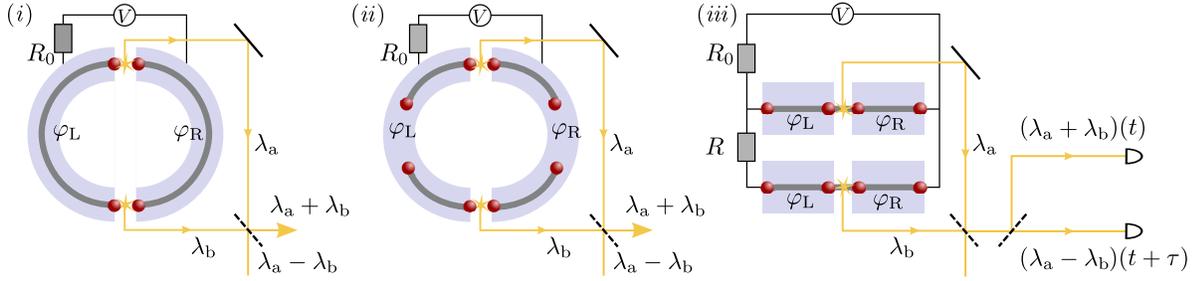}
  \caption{Three different designs for small networks with two Josephson
  junctions.  In (\emph{i}), a single semiconducting (circular) nanowire is
  lying on top two bulk superconductors realizing a d.c.-SQUID geometry
  interrupted by two Josephson junctions.  In the setup (\emph{ii}), the two
  fractional Josephson junctions in the d.c.-SQUID are formed by two distinct
  nanowires such that there are eight Majorana zero modes in total.  In
  (\emph{iii}), the two fractional Josephson junctions are two independent
  setups of figure~1 biased at the same voltage $V$.  Each Josephson junction
  is surrounded by a microcavity which emits coherent radiation indicated by
  beams of radiation $\lambda_{\rm a}$ and $\lambda_{\rm b}$.  These radiation
  fields are subsequently lead to interfere by a semitransparent mirror
  (dashed line).  In the main text, we compare the expected coherence of the
  radiation emitted in these different setups with each other.}
  \label{Fig:Setups}
\end{figure}
%
\subsection{Autocorrelations and partial coherence}
Including the action of both switching mechanisms, the question arises on what
time-scale the Josephson radiation remains coherent.  In this section we want
to address the question of autocorrelations of the radiation emitted by a
single Josephson junction, whereas the correlations between different sources
are discussed in the next section.

The correlations of the radiation field can be determined from a master
equation treatment of the switching processes.  The main task of the master
equation is to evaluate the vector $\bm{p}(t)$ whose components are the
probabilities $p_\alpha(t)$ for finding the system in a particular state
$\alpha \in|\pm, \mathcal{P} \rangle$ at time $t$ given some initial state
$\bm{p}(0)$.  The master equation describing the switching events then reads
\begin{equation}\label{eq:master}
\dot {\bm{p}} = \left[-(\Gamma_{\rs F} + \Gamma_{\rs{QP}}) \openone_4
+ \Gamma_{\rs F} \sigma_1 + \frac{\Gamma_{\rs{QP}}}{2} (\openone_2 
+ \sigma_1)\tau_1 \right] \bm{p}
\end{equation}
where $\sigma_j$ ($\{\tau_j\}$) are Pauli-matrices acting on the first
(second) label of $|\pm, \mathcal{P} \rangle$.  The time evolution of the
radiation field is thus given by
\begin{eqnarray}
  \langle \Lambda(t)\rangle = 
  \sum_{\alpha} e^{-i\varphi(t)/2} \lambda_\alpha \, p_\alpha(t)
\end{eqnarray}
where the angular brackets indicate the ensemble average and we have assumed
that the radiation field is always in its stationary state $\lambda_\alpha$
which corresponds to demanding that $\Gamma \gg \Gamma_\rs{QP},
\Gamma_\rs{F}$.  In general, correlations can be expressed in terms of the
normalized first-order correlation function
\begin{equation}
g^{(1)}(\tau)= \frac{\langle \Lambda^*(t) \Lambda(t+\tau) \rangle}{\langle
|\Lambda|^2 \rangle}.
\end{equation} 
correlating the signal $\lambda(t)$ with itself after a time $\tau$.  Without
the switching processes discussed before, the autocorrelation function is one
indicating coherence with an infinite coherence time.  The situation changes
when spontaneous switches are taken into account.  Then the relative phase
factors of $\lambda$ changes randomly by $\pi$ resulting in a finite coherence
time.

Autocorrelations of each beam can be measured by performing a
Hanbury Brown-Twiss  experiment measuring intensity correlations of the
radiation field at the different times $t$ and $t+\tau$.  In fact, the first
order autocorrelation function and with that its coherence time can be
extracted from the normalized second order correlation function via
\begin{equation}
g^{(2)}	= 	\frac{\langle \Lambda^*(t) \Lambda^*(t+\tau) \Lambda(t+\tau)
\Lambda(t) \rangle}{\langle |\Lambda|^2 \rangle^2 }
				=	1 + |g^{(1)}(\tau)|^2
\end{equation}
where the last identity is valid for a classical radiation field as we are
considering \cite{Loudon:1973kx}. The solution of the master equation for
$\bm{p}$ yields in the steady state the first-order correlation function
\begin{equation}
  g^{(1)}(\tau) = \exp \left(  -\frac{i}2 \omega_{\rs J} \tau - \frac{\tau}{\tau_{\rm
c} } \right)
\end{equation}
where $\tau_{ \rm c} = (2\Gamma_{\rs{F}} + \Gamma_{\rs{QP}})^{-1}$ is the
coherence time.  The reason for the factor two of $\Gamma_\rs{F}$ compared to
$\Gamma_\rs{QP}$ is the fact that the former process leads every time to a
switch of the upper to the lower state whereas for the former process this
happens only with a probability of 50\%.

Another effect which leads to decoherence are thermal fluctuations of the bias
voltage.  Since there is a finite resistance $R_0$ associated to the circuit
that connects the ideal voltage source with the Josephson junction,
cf.~figure~\ref{Fig:Setups}, there is a Johnson-Nyquist noise characterized by
$\langle\langle V(t) V(t+\tau) \rangle\rangle= 2 R_0 k_{\rs B}T \delta(\tau)$
at any finite temperature $T$.  As the first-order correlation function is
proportional to the complex phase factor $e^{i [\varphi(t)
-\varphi(t+\tau)]/2}$ and thus depends on $V(t)$, the thermal noise leads to
decoherence.  Using the fact that the thermal noise is Gaussian, the expectation
value of the complex phase factor assumes the form
\begin{equation}
  \fl
\langle e^{i [\phi(t) -\phi(t+\tau)]/2}  \rangle	
= e^{ - i \omega_\rs{J} \tau/2} e^{- \langle\langle [\varphi(t)-
\varphi(t+\tau)]^2 \rangle\rangle /8} = \exp \left( -
\frac{i}{2} \omega_\rs{J} \tau
-   \frac{e^2 R_0}{\hbar^2}k_{\rs B} T\tau \right).
\end{equation}
Consequently, the normalized auto-correlation function is given by
\begin{equation}
  g^{(1)}(\tau) =  \exp \left[ -
\frac{i}{2} \omega_\rs{J} \tau
-  \left( \tau_{\rm c}^{-1} +
\frac{e^2 R_0}{\hbar^2}k_{\rs B} T \right)\tau \right]
\end{equation}
with a reduced coherence time due to the finite resistance $R_0$.
%
\subsection{Correlations of different sources}
So far, we have discussed the dynamics and stationary properties of the
emitted Josephson radiation of a single junction.  In this section, we want to
expand the setup and take possible coherence between different emitters into
account.  Due to the phase locking, we expect that the radiation fields are
ideally perfectly correlated with each other \cite{Recher:2010qy}.  Hence, it
should be possible to observe correlations between two radiation fields even
if these junctions are spatially-separated from each other.

Here, we want to determine the correlation between different radiation sources
for the three setups illustrated in figure~\ref{Fig:Setups}.  We will label
the properties associated with two radiation sources by a and b.  The
coherence will show up in the second-order correlation function
\begin{equation}
  g_{\rm ab}^{(2)}(\tau) = \frac{\langle \Lambda^*_{\rm a}(t) \Lambda^*_{\rm
  b}(t+\tau) \Lambda_{\rm b}(t) \Lambda_{\rm a}(t+\tau) \rangle}
  {\langle |\Lambda_{\rm a}|^2 \rangle \langle |\Lambda_{\rm b}|^2 \rangle}
\end{equation}
measuring correlations between two radiation fields $\lambda_{\rm a}$,
$\lambda_{\rm b}$.  The function $g_{\rm ab}^{(2)}(\tau)$ is part of the
intensity correlators $\langle |\Lambda_{\rm a}(t) \pm \Lambda_{\rm b}(t)|^2
\, |\Lambda_{\rm a}(t+\tau) \pm \Lambda_{\rm b}(t+\tau)|^2 \rangle $ and thus
can be measured by correlating the intensities after the beamsplitters, see
figure~\ref{Fig:Setups}~(\emph{iii}).

The first setup, we analyze is shown in
figure~\ref{Fig:Setups}~(\emph{i}). There is a single circular nanowire placed
on a superconducting ring interrupted by two Josephson junctions (d.c.-SQUID
geometry).  Each of these junctions implements a source as discussed above.
For simplicity, we assume both junctions to be equal, i.e., $g_{\rm a} =
g_{\rm b}=g$, $\Omega_{\rm a} = \Omega_{\rm b}=\Omega$, and $\Gamma_{\rm a} =
\Gamma_{\rm b}=\Gamma$.  In total there will be four Majorana fermions in the
system and the equations of motion assume the form
\begin{eqnarray}
  \dot\lambda_{\rm a}	
&=	-(i\Omega + \Gamma)\lambda_{\rm a} 
\pm i \frac{g^2}{4\hbar} \, \frac{\Re (\lambda_{\rm a} - \mathcal{P}
\lambda_{\rm b}) }{\sqrt{ \delta_{\mathcal{P}}^2 + g^2 \Re^2 (\lambda_{\rm a}  -
\mathcal{P} \lambda_{\rm b})}},  \nonumber \\
	\dot\lambda_{\rm b}
&=	-(i\Omega + \Gamma)\lambda_{\rm b} 
\pm i \frac{g^2}{4\hbar} \, \frac{ \Re (\lambda_{\rm b} - \mathcal{P} \lambda_{\rm a}) } 
	{\sqrt{ \delta_{\mathcal{P}}^2 + g^2 \Re^2 (\lambda_{\rm a}  - \mathcal{P}
  \lambda_{\rm b})}}, \label{Eq:Equations-of-Motion-Ring}
\end{eqnarray}
with $\mathcal{P}$ the total fermion parity of the system, as before.  The
stationary solutions are given by (\ref{Eq:Stationary-Points}) with the
additional constraint $\lambda_{\rm b} = -\mathcal{P} \lambda_{\rm a}$ which
correlates both field amplitudes $\lambda_{\rm a}$ and $\lambda_{\rm b}$ by
the parity constraint.  Hence the d.c.-SQUID geometry exhibits a strong
correlation between different stationary fields for arbitrary
spatial-separation.  In particular, the constraint implies that the relative
phase between $\lambda_{\rm a}$ and $\lambda_{\rm b}$ can only be changed by
changing the total fermion parity of the system, i.e., quasi-particle
poisoning.  Indeed, we obtain the result
\begin{equation}
g_{\rm ab}^{(2)}(\tau) = \exp(-2 \Gamma_{\rs{QP}} \tau).
\end{equation}
for the case (\emph{i}).  Note that this result is independent of the
resistance $R_0$ of the voltage source.\footnote{Interestingly, for case
(\emph{i}) one can also obtain a non-vanishing cross-correlation
$g_{12}^{(1)}(\tau)= \exp(-i\omega_{\rs J} \tau/2 - \tau/\tau_{\rm c})$,
because it behaves essentially like an autocorrelation according to the constraint
$\lambda_{\rm b} = -\mathcal{P} \lambda_{\rm a}$.}

The case (\emph{i}) shows very robust correlations but we expect it to be
rather challenging to realize experimentally as a circular nanowire is
required.  Therefore, we want to analyze the situation where the two emitters
are formed by two different nanowires, see
figure~\ref{Fig:Setups}~(\emph{ii}).  In contrast to the last case, there are
in total eight Majorana fermions involved in this case---four on each
nanowire.  As a result, there is no parity constraint relating the phase of
the field of one emitter to the field of the other emitter in the stationary
state.  Instead, the Hamiltonian of the wires separates into two parts $H_{\rm
a} + H_{\rm b}$ with only the common superconducting phase providing
correlations.  Thus, the stationary state of the radiation fields in the
rotating frame are individually given by (\ref{Eq:Stationary-Points}).  In the
laboratory frame, the states evolve according to $e^{-i \varphi(t)/2}
\lambda_{\rm a/b}$ with the common superconducting phase-difference
$\varphi(t)$ which is expected to lead to partial coherence of the sources.
Indeed, calculating the correlation function for this case, we obtain
\begin{equation}\label{eq:ii}
g_{\rm ab}^{(2)}(\tau) = \exp[-2(2\Gamma_{\rs F} + \Gamma_{\rs{QP}})\tau]
\end{equation}
which is decaying with the typical time scale $\tau_{\rm c}$ of the
spontaneous switching events.  As above, the result is independent of the
resistance $R_0$.

The last case (\emph{iii}) differs from (\emph{ii}) by the fact that the
radiation-emitting junctions do not share common superconductors but only a
common voltage source.  In this case, the phase-difference $\varphi_{\rm a}$
will differ from $\varphi_{\rm b}$ which leads to decoherence of the two
radiation fields evolving according to $e^{-i \varphi_{\rm a/b}(t)/2}
\lambda_{\rm a/b}$ in the laboratory frame.  In fact, the diffusion of the
difference $\varphi_{\rm a}- \varphi_{\rm b}$ will be governed by the
resistance $R$, cf.\ figure~\ref{Fig:Setups}~(\emph{iii}).  Because of that
the second-order correlation function
\begin{equation}
	g_{\rm ab}^{(2)}(\tau)  = \exp\left[-2(2\Gamma_{\rs F} + \Gamma_{\rs{QP}}) \tau
  - \frac{e^2 R}{\hbar^2} k_{\rs B}T \tau \right].
\end{equation}
shows an additional decay compared to (\ref{eq:ii}).
%
\section{Conclusions}
In this work, we have analyzed the possibility of coupling Majorana fermions
to electromagnetic fields.  We have shown that partially coherent Josephson
radiation is emitted at half of the Josephson frequency.  The coherence of the
radiation is limited due to rare spontaneous switches of the relative phase
flipping randomly between values $0,\pi$.  Due to the coupling of the Majorana
zero modes to the superconductor, there are only two phases differences
allowed.  It leads to a fixed phase of the radiation field different from
conventional lasers where the optical phase is slowly diffusing
\cite{Scully:1967fk}.  Even though the phases are in principle locked,
decoherence of the radiation is induced by spontaneous switches as well as by
frequency fluctuations.  We have analyzed the correlation between two emitters
which are spatially-separated but share the same superconductors.  We have
discussed the effect of the fermionic parity constraint in a d.c.-SQUID
geometry as well as the pinning of the Josephson frequency in terms of a
second order correlator and suggested a possible way to experimentally obtain
this information.  The interaction of Majorana fermions with radiation could
potentially be used for addressing and manipulating Majorana states beyond
driving since the radiation field carries all the information about the state
of the nanowire.

\ack
We acknowledge fruitful discussions with F.\ Konschelle as well as financial
support from the Alexander von Humboldt foundation.
%
\appendix
%
%
\section{Dipole operator for Majorana fermions}\label{Ap::BdG-Dipole}
The matrix elements of the electric dipole operator in the Majorana ground
state manifold are solutions of the Bogoliubov-de Gennes equation,
$\mathcal{H}_{\rm BdG} \,\xi_n(\bm{r}) = E_n \,\xi_n(\bm{r})$ with the
Nambu-spinor $\xi_n(\bm{r}) = (u_n(\bm{r}) , v_n(\bm{r}))^{\rm t}$
\cite{Tinkham:1996fk}.  From these solutions $\{ \xi_n \}$ the Bogoliubov
quasiparticle operators can be obtained obtained via
\begin{equation}
	\beta_{n} 			= \int [u^*_{E_n}(\bm{r}) \psi(\bm{r}) + v^*_{E_n}(\bm{r})
  \psi^\dag(\bm{r})] \, d^3 r \label{Eq:Bogoliubov-Operators},
\end{equation}
with $\beta_{n},\beta_{n}^\dagger$ obeying the canonical anticommutation
relations, $\{\beta_m, \beta_n^\dag \} = \delta_{mn}$.  Every BdG-Hamiltonian
carries a build-in particle hole symmetry which is represented by an operator
$\Xi= \tau_x \mathcal{K}$ that is anti commuting with the BdG Hamiltonian $\{
\mathcal{H}_{\rm BdG} , \Xi \}=0$.  Here $\tau_x$ denotes the Pauli $x$ matrix
in Nambu space and $\mathcal{K}$ is the complex conjugation.  The operator
$\Xi$ maps a solution of the BdG equation to its particle-hole reversed
partner $\xi_{-n} = \Xi \, \xi_{n}$, which is again a solution of the BdG
equation having eigenvalue $-E_n$.  In particular, the eigenspace related to
solutions at zero energy $\{ \xi_{\mu} | \; \mathcal{H}_{\rm BdG} \xi_\mu = 0
\}$ needs to contain an even number of solutions.  Given a zero energy
solution $\xi_\mu$, one can choose
a different basis which is the eigenbasis of the
particle-hole operator via $\zeta_{\mu} = \xi_\mu +  \Xi\xi$ and
$\bar{\zeta}_{\mu} = i (\Xi\xi_{\mu} - \xi_{\mu})$.  The new spinors are
invariant with respect to electron-hole inversion $\Xi \zeta_{\mu} =
\zeta_{\mu}$ which constitutes a condition for Majorana fermions.  The new
spinors are given by $\zeta_\mu = (w_\mu, w_\mu^*), \bar{\zeta}_\mu =
(\bar{w}_\mu, \bar{w}^*_\mu)$ with components $w_\mu(\bm{r}) = u_\mu(\bm{r}) +
v_\mu^\ast(\bm{r})$ and $\bar{w} = i u_\mu(\bm{r}) - iv_\mu^*(\bm{r})$.  In
the language of second quantization one gets a new set of \emph{Hermitian}
operators
\begin{eqnarray}
	\gamma_{\mu}			&=	\beta^\dagger_{\mu} + \beta^{\pdag}_{\mu},\qquad
	\bar{\gamma}_{\mu}	=	i (\beta^\dagger_{\mu} - \beta^{\pdag}_{\mu}),
\end{eqnarray}
fulfilling the algebra of Majorana bound states $\{ \gamma_\mu, \gamma_\nu \}=
2\delta_{\mu \nu}$.  The existence of these zero energy modes is guaranteed by
the topological structure of the BdG Hamiltonian $\mathcal{H}_{\rm BdG}(k)$ in
presence of (effective) \emph{p}-wave pairing potentials.  Using the reversed
Bogoliubov transformation (\ref{Eq:Bogoliubov-Operators}),
\begin{equation}
	\psi(\bm{r}) 				= \sum_m [u_m(\bm{r}) \beta_m + v^*_m(\bm{r})
  \beta_m^\dagger ],
\end{equation}
it is a straightforward task to represent a second quantized operator in terms
of Bogoliubov quasi-particles.  By restricting the general expression on the
subspace of zero energy solutions only, any observable $O$ can be expressed in
terms of Majorana operators as follows
\begin{eqnarray}
	O	&= \int 	\psi^\dagger(\bm{r}') O(\bm{r}',\bm{r}) \psi(\bm{r}') \, d^3 r'
  d^3 r \nonumber\\
		&=	\frac{1}{4} \sum_{\mu,\nu} \left[ \int w^\ast_\mu(\bm{r}')
    O(\bm{r}',\bm{r}) w_\nu(\bm{r}) \, d^3 r' d^3 r \right] \gamma_\mu
    \gamma_\nu =
	\frac{i}{4} \sum_{\mu, \nu} \mathcal{O}_{\mu\nu} \gamma_{\mu} \gamma_{\nu}, 
\end{eqnarray}
where the sum runs over all Majorana fermions $\gamma_\mu, \bar{\gamma}_\mu$
in the last two lines.  As the coefficient in front of the Majorana operators
is a Hermitian scalar product with respect to the Majorana wave functions
$w_\mu(\bm{r}')$ and $w_\nu(\bm{r})$.  This matrix element can only have a
finite value where two Majorana wave functions have a considerable overlap.
Because the Majorana wave function is spatially localized at the ends of the
nanowire finite matrix elements can only appear at Josephson junctions where
two Majorana modes are close together.  Note, that the matrix elements are 
be antisymmetric, $\mathcal{O}_{\mu \nu} = - \mathcal{O}_{\nu \mu}$, in the
Majorana basis.

Matrix elements between Majorana operators located at two different sides of a
Josephson junction acquire a non-trivial phase dependency.  This can be
derived by performing a gauge transformation of the BdG Hamiltonian.  From
general principles, it is clear that a non-trivial phase difference $\varphi
\neq 0$ cannot  be gauged away.  Therefore all matrix elements
connecting degrees of freedom on both sides of the Josephson junction acquire
phase factors $e^{\pm i \varphi/2}$ whereas all others become independent of
the superconducting phases.  In particular, if the operator $O$ connects two
Majorana fermions $\gamma_\mu, \gamma_\nu$ across a Josephson junction we
obtain the representation
\begin{eqnarray}
 O	&=		\frac{i}{4} e^{i \varphi/2} \mathcal{O}_{\mu \nu} \gamma_{\mu}
 \gamma_{\nu} + \frac{i}{4} e^{-i \varphi/2}  \mathcal{O}_{\nu \mu}
 \gamma_{\nu} \gamma_{\mu} 
	= 	\frac{i}{2} \cos(\varphi/2) \mathcal{O}_{\mu \nu} \gamma_\mu \gamma_\nu,
\end{eqnarray}
where the dependence on the superconducting phase keeps track of the charge
that is transported across the junction.  Accordingly, for the electrical
dipole operator we get the following representation in terms of Majorana
operators
\begin{equation}
	\bm{d} 	= - e \int \psi^\dagger(\bm{r}) \, \bm{r} \psi(\bm{r}) \, d^3r		
			= \frac{i}{2} \sum_{\mu < \nu} \cos(\varphi/2) 
      \bm{d}_{\mu \nu} \gamma_{\mu} \gamma_{\nu}
\end{equation}
where the matrix elements of the dipole operator are given by 
\begin{equation}
	\bm{d}_{\mu \nu} = -e \int \bm{r} \, w^*_{\mu}(\bm{r}) w_{\nu}(\bm{r}) \, d^3r  .
\end{equation}
Note again that these quantities are only non-zero if the Majorana wave
functions have some overlap.
%
\section{Influence of counter-rotating terms}\label{Ap:Counter-Rotating-Terms}
In order to study the effect of the counter-rotating terms that have been
neglected in (\ref{Eq:RWA-Hamiltonian}) the non-rotating stationary solutions
are put into the original Hamiltonian (\ref{Eq:Model-Hamiltonian})
substituting operator $a$ by $\exp(-i \varphi(t)/2)\lambda$.  The resulting
Hamiltonian is given by
\begin{eqnarray}
  \fl
H 	&=	\frac{\varepsilon_{\rs L} + \varepsilon_{\rs R}}{2} \openone_2 + \frac{\delta_{\mathcal{P}}}{2} \sigma_3 - \frac{g |\lambda|}{2} \cos(\chi) \sigma_1 
 	+ \frac{g |\lambda|}{2} \cos(\omega_{\rs J} t + \varphi(0) - \chi) \sigma_1,
\end{eqnarray}
where the last term represents fast oscillating counter-rotating terms, that
have been neglected while going from (\ref{Eq:Model-Hamiltonian}) to
(\ref{Eq:RWA-Hamiltonian}).  Furthermore, $\chi = \arctan(\Gamma/\Omega)$
is the phase shift due to the cavity and $\sigma_1 = i \gamma_2 \gamma_3,
\sigma_3 = -i\gamma_1 \gamma_2 $ are Pauli matrices.  This is a time-dependent
problem with driving frequency $\omega_{\rs J}$.  After interchanging $\sigma_1$
and $\sigma_3$, we make the ansatz \cite{Nazarov:2009uq}
\begin{equation}
	|\phi \rangle = \exp\left( i \frac{g |\lambda|}{2\hbar} \int\limits_0^t \cos(\omega_{\rs J} t + \varphi(0) - \chi) dt\,  \sigma_3 \right) | \tilde{\phi} \rangle
\end{equation}
for the wave function $|\phi\rangle$.  This leads to the Schr\"odinger
equation for $|\tilde{\phi}\rangle$
\begin{equation}\label{Eq:Effective-Hamiltonian-CRT}
	i \hbar \partial_t 
  \left(\begin{array} {c}\tilde{\phi}_1 \\ \tilde{\phi}_2 \end{array}\right) 
    = \frac{1}{2} 
    \left(\begin{array}{cc} 	- g |\lambda| \cos{\chi} 			&	e^{i A(t)} \delta_{\mathcal{P}} \\   
										e^{-i A(t)} \delta_{\mathcal{P}}	&	g |\lambda| \cos{\chi}
                  \end{array} \right)
  \left(\begin{array}{c} \tilde{\phi}_1 \\ \tilde{\phi}_2 \end{array}\right)
\end{equation}
with the time-dependent expression $A(t) = \frac{g |\lambda|}{2\hbar
\omega_{\rs J}} [ \sin(\omega_{\rs J}t + \varphi(0) - \chi) - \sin(\varphi(0)
- \chi)]$.  Since the driving frequency is much larger than the largest energy
scale in (\ref{Eq:Effective-Hamiltonian-CRT}), $g |\lambda| \ll \hbar
\omega_{\rs J}$ it is reasonable to perform a time average over one period $T=
\omega_{\rs J}/ 2\pi$.  From the resulting time-averaged Schr\"odinger
equation the probability for a state inversion flip is obtained to
\begin{eqnarray}
P_{+\rightarrow -}(t) &= | \langle \phi_- | \phi_+(t) \rangle |^2 
\approx  \left(\frac{\delta_{\mathcal{P}}}{g\,\Re\lambda} \right)^2
\sin^2\left(\frac{g\,|\Re\lambda|}{2\hbar}\, t \right).
\end{eqnarray}
The probability to flip from the upper to the lower state is apparently
oscillating in a coherent manner with period $4\pi\hbar /g\,|\Re\lambda| \ll
\Gamma^{-1}$.
%
%
%
%

\end{document}